%% file: main.tex
\documentclass[lettersize,journal]{IEEEtran}
\usepackage{amsmath,amsfonts}
\usepackage{array}
\usepackage[caption=false,font=normalsize,labelfont=sf,textfont=sf]{subfig}
\usepackage{textcomp}
\usepackage{stfloats}
\usepackage{url}
\usepackage{verbatim}
\usepackage{graphicx}
\usepackage{cite}
\usepackage{xspace}

\usepackage{subcaption}
\usepackage{graphicx}
\usepackage[linesnumbered]{algorithm2e}
\usepackage[dvipsnames]{xcolor}
\usepackage{tikz}

\usepackage{multirow}
\usepackage{enumitem}
	
\usepackage{soul}

\usepackage{array}

\usepackage{tikz}
\newcommand*\emptycirc[1][1ex]{\tikz\draw (0,0) circle (#1);} 
\newcommand*\halfcirc[1][1ex]{%
  \begin{tikzpicture}
  \draw[fill] (0,0)-- (90:#1) arc (90:270:#1) -- cycle ;
  \draw (0,0) circle (#1);
  \end{tikzpicture}}
\newcommand*\fullcirc[1][1ex]{\tikz\fill (0,0) circle (#1);} 

\usepackage{tablefootnote}

\usepackage{cleveref} 
\newtheorem{theorem}{Theorem}
\newtheorem{lemm}[theorem]{\textbf{Lesson}}
\usepackage{mdframed}

\begin{document}

\title{To See or Not to See - Fingerprinting Devices in Adversarial Environments Amid Advanced Machine Learning}

\author{Justin Feng \IEEEmembership{Student, IEEE}, 
Amirmohammad Haddad \IEEEmembership{Student, IEEE}, Nader Sehatbakhsh \IEEEmembership{Member, IEEE}
}

\markboth{}%
{Shell \MakeLowercase{\textit{et al.}}: SoK: To See or Not to See - Fingerprinting Devices in Adversarial Environments Amid Advanced Machine Learning}


\maketitle

\begin{abstract}
The increasing use of the Internet of Things raises security concerns. To address this, device fingerprinting is often employed to authenticate devices, detect adversaries, and identify eavesdroppers in an environment. This requires the ability to discern between legitimate and malicious devices which is achieved by analyzing the unique physical and/or operational characteristics of IoT devices. In the era of the latest progress in machine learning, particularly generative models, it is crucial to methodically examine the current studies in device fingerprinting. This involves explaining their approaches and underscoring their limitations when faced with adversaries armed with these ML tools. To systematically analyze existing methods, we propose a generic, yet simplified, model for device fingerprinting. Additionally, we thoroughly investigate existing methods to authenticate devices and detect eavesdropping, using our proposed model. We further study trends and similarities between works in authentication and eavesdropping detection and present the existing threats and attacks in these domains. Finally, we discuss future directions in fingerprinting based on these trends to develop more secure IoT fingerprinting schemes.
\end{abstract}

\begin{IEEEkeywords}
Embedded Devices, Secure Communications, and Mobile and Ubiquitous Systems.
\end{IEEEkeywords}

\maketitle

\section{Introduction}
\label{sec:intro}
\input{Sections/introduction}

\section{Background}
\label{sec:back}
\input{Sections/background}

\section{Device Authentication}
\label{sec:fing}
\input{Sections/fingerprinting}

\section{Eavesdropping Detection}
\label{sec:hidden}
\input{Sections/hidden_devices}

\section{Existing Adversarial Attacks}
\label{sec:spoof}
\input{Sections/spoofing}

\section{Future Attack Scenarios}
\label{sec:atk}
\input{Sections/attack}

\section{Discussions}
\label{sec:disc}
\input{Sections/discussions}

\section{Conclusions}
\label{sec:conc}
\input{Sections/conclusions}

\bibliographystyle{IEEEtran}
\bibliography{references}

\end{document}

%% file: Sections/introduction.tex
\begin{figure}[t]
    \centering
    \includegraphics[width=1\columnwidth]{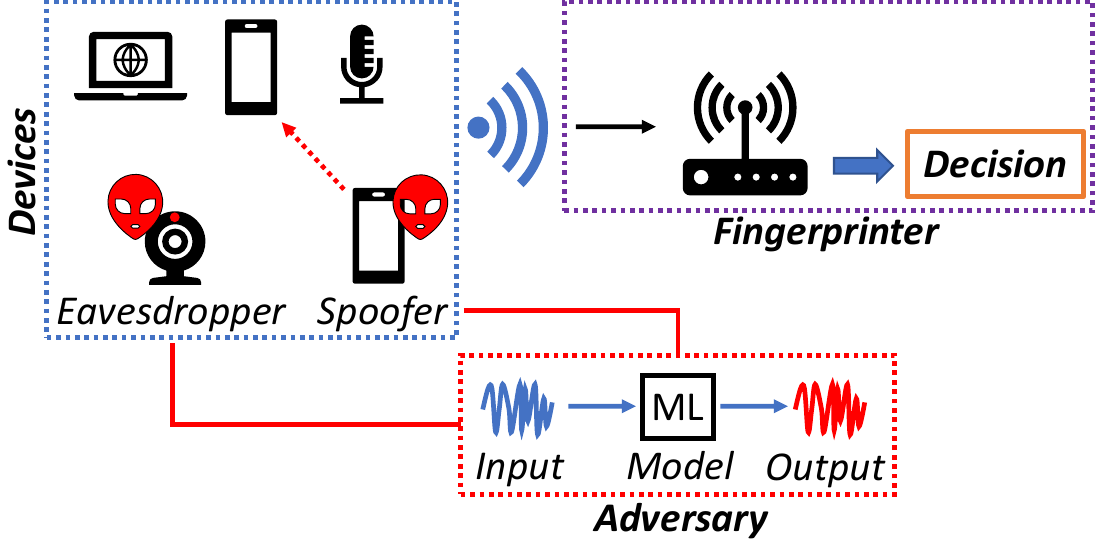}
    \caption{
    In an adversarial fingerprinting setting, the adversary may target the system by designing a spoofing method to bypass the authentication and/or by installing a hidden device (eavesdropper) to steal information.}
    \label{fig:physical_setup}
\end{figure}

The Internet of Things (IoT) is becoming increasingly commonplace in our lives, bringing with it numerous benefits and opportunities such as smart homes, smart manufacturing, and wearable and mobile computing. However, the integration of IoT devices in these environments and applications raises the stakes of potential security attacks. These attacks could have devastating and far-reaching consequences for affected systems, not just for the individual user but for an entire network, leading to catastrophic outcomes~\cite{joo2020hold, givehchian2022evaluating, singh2021always, chaman2018ghostbuster, bao2021threat}. Therefore, it is essential to take the necessary steps to ensure that IoT systems are secure and protected from malicious actors.

Practical solutions for protecting IoT devices and networks against attacks rely on \textit{authentication}. The authentication process often involves leveraging cryptographic primitives such as message authentication codes (MACs). These cryptographic-based solutions provide provable security guarantees, however, they impose large overheads on the IoT system. This is an important concern since IoT devices are typically resource-constrained, and in many cases, solutions with large power and/or storage overhead are not practical and/or feasible. Moreover, cryptographic-based methods are subject to various side-channel and/or physical attacks~\cite{zhao2018fpga, genkin2015get, lipp2021platypus, guri2015gsmem, camurati2020understanding}. 

An alternative approach for device authentication that is lightweight yet still effective relies on \textbf{\textit{fingerprinting}}, where various unique characteristics of the device (e.g., its RF signal, physical radiation, etc.) are leveraged to identify the device.

Apart from the concern of authentication, the prevalence of IoT devices has created another new security threat: \textbf{\textit{eavesdropping}}. In this scenario, the device itself is not the target. Instead, attackers misuse IoT devices to launch \textit{privacy attacks}, such as creating hidden voice recorders, network packet sniffers, or hidden cameras. 

Both of these issues have a significant common trait: the need to identify the IoT device by detecting and analyzing the \textit{physical signals} it generates. In the case of fingerprinting, this identification results in authentication. In eavesdropping, it aims for privacy protection. Crucially, they both have a shared \textbf{\textit{vulnerability}}: susceptibility to being deceived by an adversary who can manipulate the detection mechanism. An overview of these security threats is shown in Figure~\ref{fig:physical_setup}.

With the recent strides in machine learning, particularly in \textbf{\textit{generative models}} like GANs, LLMs, and diffusion models~\cite{li2021gnss, ayanoglu2022machine, creswell2018generative, chen2021evaluating, wei2022emergent, croitoru2023diffusion, yang2022diffusion}, it is essential to extensively examine the strength of current solutions against these types of attacks. This paper aims to expand on prior fingerprinting research surveys~\cite{xu2015device, soltanieh2020review, liu2021machine, xie2020survey}, delving into a systematic exploration of attack and defense trends for IoT device fingerprinting and eavesdropping. Furthermore, it delves into future attack trends in light of the advancements in machine learning.

\begin{figure}[t]
    \centering
    \includegraphics[width=1\columnwidth]{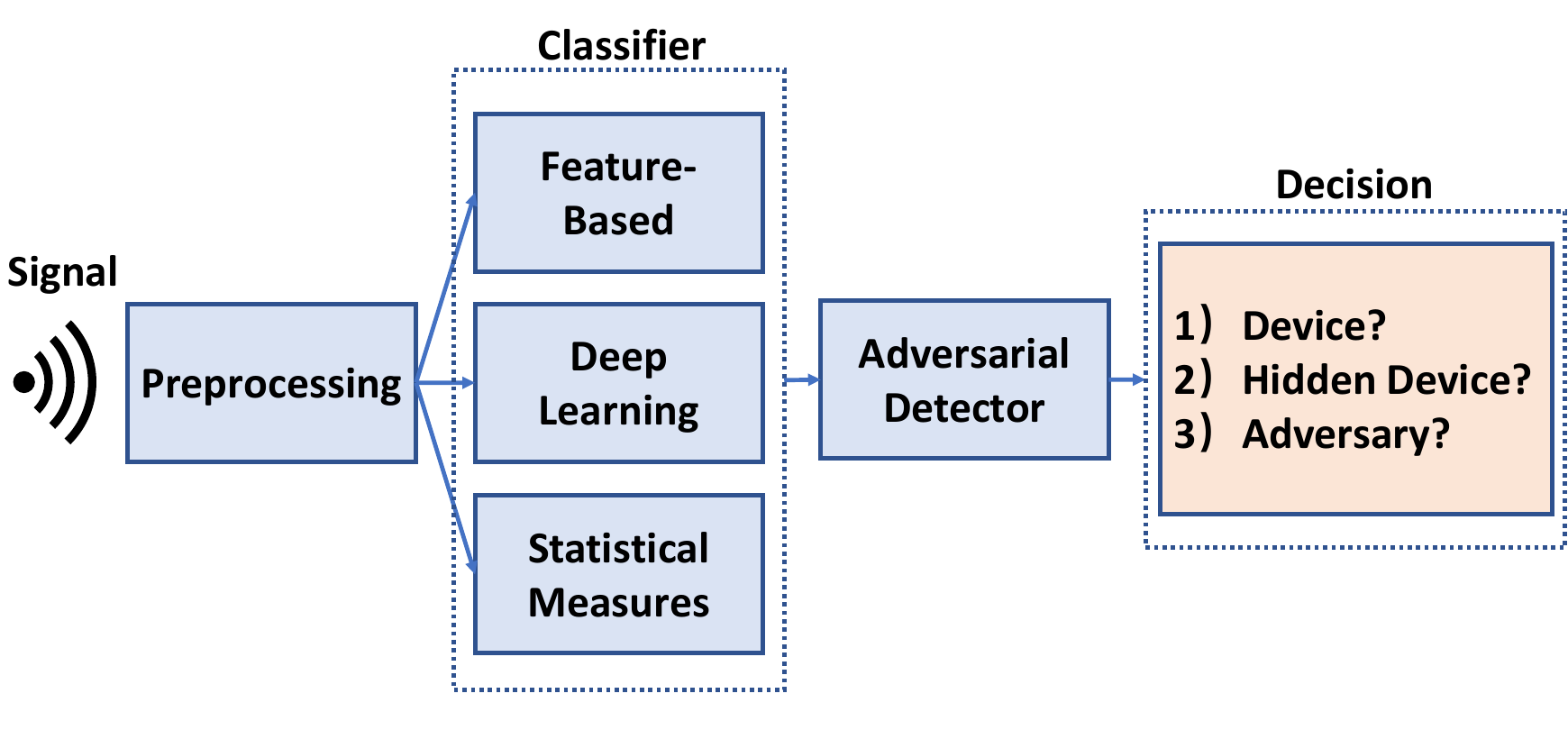}
    \caption{Overall diagram of fingerprinting systems. The system receives a \textit{signal} as input and \textit{preprocesses} it. This data is then sent to a \textit{classifier}. An \textit{adversarial detector} detects an attack. Finally, we have a \textit{Decision} which tells us 1) which device was authenticated, 2) was a hidden device detected, and 3) was an adversary found?}
    \label{fig:overall_diagram}
    \vspace{-5pt}
\end{figure}

To provide a comprehensive study, we first systematize the fingerprinting process using a formal model for fingerprinting and the adversary (\S\ref{sec:back}). To study the future trends, especially the impact of advancements in machine learning on fingerprinting robustness, we first thoroughly explain the existing work on device identification and eavesdropping (\S\ref{sec:fing} and \S\ref{sec:hidden}) and then describe the future attack strategies (\S\ref{sec:atk}). We conclude the paper with some discussion points (\S\ref{sec:disc}) and conclusions (\S\ref{sec:conc}). 

\vspace{3pt}
\noindent \textbf{Contributions.} Our primary contributions relate to a set of formalisms for expressing the security of fingerprinting systems more effectively. Specifically: 
\begin{itemize}[leftmargin=10pt]
    \item \textbf{Fingerprinting Model}: We develop an equation that systematizes fingerprinting systems and the inputs, methods, and outputs of these systems. The hope is to enable designers to build more robust and usable authentication schemes.
    \item \textbf{In-depth analysis}: We discuss various metrics of performance for fingerprinting systems such as range, scalability, and spoofing detection. By formalizing these factors, we hope that future systems will be designed with these factors in mind in order to increase performance.
    \item \textbf{Future Trends}: We comment on design decisions of existing works (experiments, scenarios, etc.) and how future works can improve upon these choices, especially in light of new machine-learning techniques.
\end{itemize}

%% file: Sections/background.tex
\subsection{Fingerprinting Methodology}\label{subsection:back_method}
To systematize the fingerprinting process, we develop a generic yet simple model to formalize the problem, shown in Figure \ref{fig:overall_diagram} and in Equation \ref{eq:formula} which is described in the following.

\begin{figure}[t]
    \centering
    \includegraphics[width=1\columnwidth]{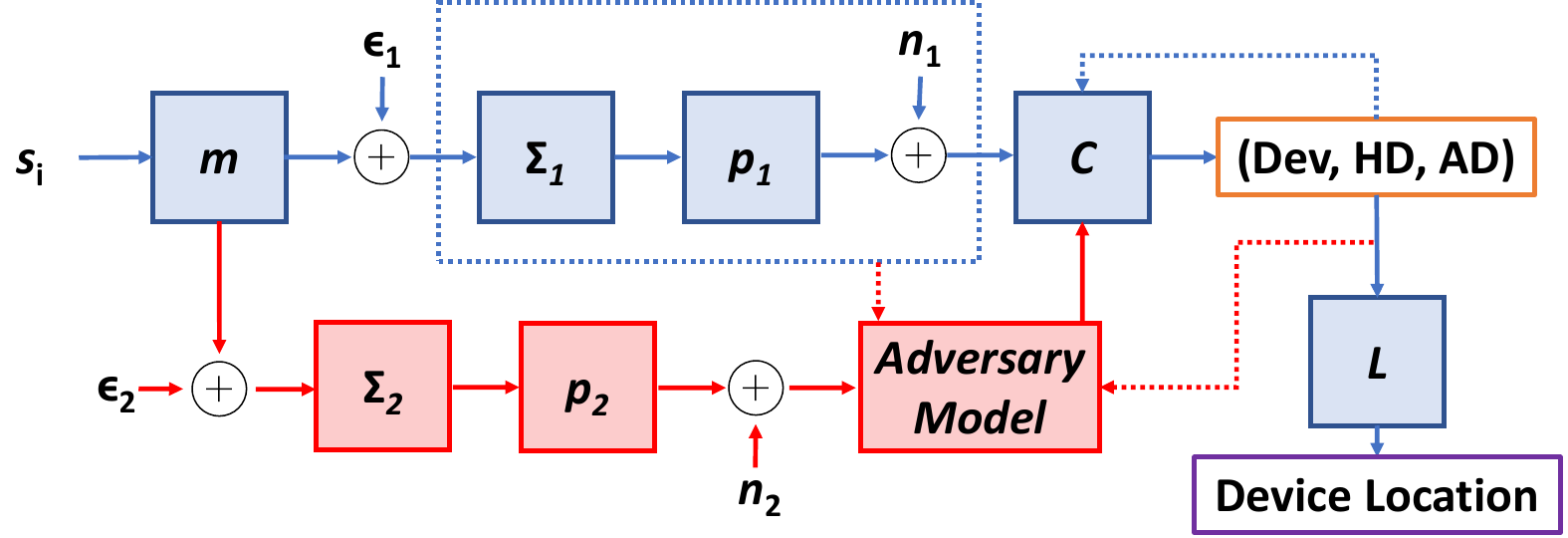}
    \caption{Systematized view of fingerprinting systems. Our input takes on a certain modulation \textit{m} depending on the system. For each sample, an error term $\epsilon$ is added. We then average our samples (\textit{$\Sigma$}) and preprocess our data with \textit{p}. During the preprocessing stage, we take on another error term $n$. Finally, we classify our device with \textit{C} and attempt to localize our device with \textit{L}. An adversary can utilize its own processing and model to try to fool the classifier \textit{C}, receiving different types of input depending on the threat model.}
    \label{fig:formula_figure}
    \vspace{-3pt}
\end{figure}

The \textit{Signal} received from a device first goes through a \textit{Preprocessing} stage where various signal processing techniques such as averaging and smoothing are applied. A \textit{Classifier} block is then used where we identify the device based on a model, which can range from \textit{Statistical Measures} to \textit{Feature-Based} methods to \textit{Deep Learning} models. Optionally, an \textit{Adversarial Detector} block could be used to detect whether this signal is genuine or malicious. This block is flexible in that it can occur before or after the \textit{Classifier} block. The goal is to identify features that make the sample fundamentally different from the authenticated devices. Finally, we have a \textit{Decision} block where we output three items: \textbf{1)} Which (authenticated) device is this? \textbf{2)} Is there a hidden/eavesdropping device detected? \textbf{3)} Is an adversary (spoofing) detected?

\subsection{Fingerprinting Model}\label{subsection:back_model}
To more concretely formalize this problem, Figure \ref{fig:formula_figure} describes a set of functions we apply to our input to receive our decision. We can describe our system formally as an equation shown below:
\begin{equation} \label{eq:formula}
(Dev, HD, AD, Loc)= L(C(p(\frac{\sum_{1}^{i} m(s_i)+\epsilon}{i})+n)),
\end{equation}
where the output \textit{(Dev, HD, AD, Loc)} is a \textit{list} showing: the identity of the authenticated device (\textit{Dev}), indicating whether a \textit{hidden device} (\textit{HD}) was discovered or not, whether an \textit{adversarial device} (\textit{AD}) was detected or not, and (if applicable) the \textit{location} (\textit{Loc}) of the device. \textit{L} is a localization function that finds the device in the room. \textit{C} is a classification function that authenticates a device, detects a hidden device, or detects an adversarial device. \textit{p} is a preprocessing function that prepares the data for classification. $n$ is a constant that is a built-in error when preprocessing data. 

Studying the preprocessing phase in greater detail, the process involves averaging data \(s_i\) concerning a particular modulation \textit{m} across \(i\) samples, each associated with a specific error \(\epsilon\). Moreover, this method averages data to minimize the impact of unwanted noise and artifacts. The addition of the error term \(\epsilon\) becomes crucial to consider the expected amount of noise in any given input. The equation is displayed below:
\begin{equation} \label{eq:formula_sum}
\frac{\sum_{1}^{i} m(s_i)+\epsilon}{i}.
\end{equation}

Our system can be expanded to include an adversary model. This adversary collects a signal from a trusted transmitter, applies their own processing algorithm, and trains their own model, which uses data from the fingerprinting pipeline as input to generate adversarial data. The data utilized depends on the level of access the adversary has to the fingerprinter and model. The goal is then to deceive the classifier \textit{C}. In Figure ~\ref{fig:formula_figure}, we differentiate between the fingerprinter and adversary paths using both color and subscripts. For instance, blue \textit{$p_1$} represents the fingerprinter's processing block, while red \textit{$p_2$} represents the adversary's processing block.

Using the formula described above, in the next two sections, we systematically map the existing work in device identification (Section \ref{sec:fing}) and eavesdropping detection (Section \ref{sec:hidden}) to our model and highlight the differences and similarities. We then focus on future attacks strategies in Section \ref{sec:atk}. 

%% file: Sections/fingerprinting.tex
\begin{table*}[]
    \centering
    \caption{This is a systematization of fingerprinting works. For each modality (RF, network, and EM), there are various preprocessing tools and characteristics that are leveraged in order to classify devices. Each modality has a certain level of performance, measured by range, scalability, and whether spoofing was considered or not.}
    \begin{tabular}{|>{\centering\arraybackslash}m{1cm}|>{\centering\arraybackslash}m{1cm}|>{\centering\arraybackslash}m{1cm}|>{\centering\arraybackslash}m{3cm}|>{\centering\arraybackslash}m{4cm}|>{\centering\arraybackslash}m{1cm}|c|c|c|>{\raggedright\arraybackslash}m{2cm}|}
    \hline
    \multicolumn{2}{|c|}{\textbf{Modality}}&\textbf{Input}&\textbf{Preprocessing}&\textbf{Characteristic}&\rotatebox{90}{\textbf{Classifier}}&\rotatebox[origin=c]{90}{\textbf{Range}}&\rotatebox[origin=c]{90}{\textbf{Scalability}}&\rotatebox[origin=c]{90}{\textbf{Spoofing?}}
    &\textbf{Papers}
    \\
    \hline

    \multirow{18}{*}{RF} & WiFi & Packet & Averaging, Filtering, Slicing, Partial Equalization & CFO, I/Q Offset/Imbalance, Amplitudes & Stats., F.B., D.L. & \halfcirc & \fullcirc & \halfcirc &\cite{yu2020you, hanna2020open, jian2021radio, al2020exposing, jian2020deep, sankhe2019no, restuccia2019deepradioid, brik2008wireless, hua2018accurate, azarmehr2017wireless, agadakos2020chameleons}\\
    \cline{2-3}
    \cline{4-10}
     & BLE & Packet & Estimation, GATT & CFO and IQ Offset, GATT & Stats. & \halfcirc & \fullcirc & \emptycirc & \cite{givehchian2022evaluating, celosia2019fingerprinting}\\
    \cline{2-10}
    & ZigBee & Packet & Synchronization, Offset Calculation, DRA & DCTF, CFO, ROI & F.B., D.L. & \halfcirc & \emptycirc & \halfcirc & \cite{peng2018design, merchant2018deep, yu2019robust, merchant2019enhanced, bihl2016feature}\\
    \cline{2-10}
    & LoRa & Packet & Decoding, Synchronization, Normalization & Tx Imperfections, IQ Features & F.B., D.L. & \halfcirc & \halfcirc & \halfcirc & \cite{shen2022towards, al2021deeplora, zhang2021radio, das2018deep}\\
    \cline{2-10}
    & RFID & RFID Signal & Resolution Alignment, Sequencing, EPC Identification & Persistence Time, Normalization, Tag Imperfection, Peak Frequency & Stats., F.B., D.L. & \emptycirc & \fullcirc & \halfcirc & \cite{pan2022rf, chen2020eingerprint, joo2020hold}\\
    \cline{2-10}
    & QPSK & Symbols & DFT, Averaging & Power Amplifier Nonlinearity & D.L. & \halfcirc & \emptycirc & \emptycirc & \cite{ hanna2019deep}\\
    \cline{2-10}
    & WiMAX & Near-Transient Response & Filtering & Standard Deviation, Variance, Skewness, Kurtosis & F.B. & \fullcirc & \emptycirc & \fullcirc & \cite{reising2020radio}\\
    \cline{2-10}
    & Drones & I/Q Samples & Noise Reduction, Trace Segmentation & Parasitic Response, & D.L. & \emptycirc & \emptycirc & \fullcirc & \cite{ li2022reliable, soltani2020rf}\\
    \cline{2-10}
    \hline

    \multicolumn{2}{|c|}{Network} & %
    Packet Flows & Traffic Parsing and Collection of Data & Timing, Clock Offsets, RSS, Flow Volume & Stats., F.B. & \halfcirc & \emptycirc & \halfcirc & \cite{perdisci2020iotfinder, kohno2005remote, charyyev2020locality, miettinen2017iot, sivanathan2018classifying, xu2015device}\\
    \hline

    \multicolumn{2}{|c|}{EM} & %
    I/Q Samples & Averaging, Folding, FFT, STFT, Normalization & Clock, IC signature, Magnetic Induction, Ambient EM & Stats., F.B., D.L. & \emptycirc & \emptycirc & \emptycirc & \cite{lee2022aerokey, cheng2019demicpu, mariano2019classification, yang2016id, ahmed2018authentication, ahmed2017radiated, shen2022electromagnetic, ibrahim2020magneto, feng2023fingerprinting}\\
    \hline

    \end{tabular}
    \smallskip
    
    \textbf{Range} - \emptycirc \hspace{1mm}$<$ 10 m, \halfcirc \hspace{1mm} 10-50 m, \fullcirc \hspace{1mm}$>$ 50 m\\
    \textbf{Scalability} - \emptycirc \hspace{1mm} 10s of devices, \halfcirc \hspace{1mm} 100s of devices, \fullcirc \hspace{1mm} 1000s of devices\\
    \textbf{Spoofing?} - \emptycirc \hspace{1mm}Rarely/Never, \halfcirc \hspace{1mm} Sometimes, \fullcirc \hspace{1mm}Most of the time/Yes \\
    \label{tab:fingerprinting}
    
\end{table*}

A popular solution for device authentication relies on device \textit{fingerprinting} where physical and/or operational attributes of the device are utilized to securely authenticate the device. The key idea is that these characteristics are device-dependent and are not easily replicated, hence they can be used as unique and secure features for authentication.  

We categorize existing fingerprinting methods into three main groups: RF, network-based, and electromagnetic fingerprinting. We then describe the five fundamental steps or phases shown in Figure~\ref{fig:overall_diagram} and \ref{fig:formula_figure} for each group. These steps are input/signal ($s_i$, $m$), preprocessing ($\Sigma$, $p$), feature extraction, classification ($C$), and performance. The summary of our analysis is shown in Table~\ref{tab:fingerprinting}. 

\subsection{RF Fingerprinting}
The goal of RF Fingerprinting is to identify devices by investigating their specific RF features. For each step in fingeprinting, proper methods and design choices should be used based on the type of transmitters and receivers available.

\vspace{3pt}
\noindent
\textbf{Input.} For RF Fingerprinting systems, data can be received as an RF signal~\cite{chen2020eingerprint,joo2020hold, pan2022rf}, I/Q samples~\cite{shen2022towards,zhang2021radio,jian2021radio,soltani2020rf,al2020exposing,merchant2018deep,sankhe2019no}, and sequences of packets~\cite{yu2020you,hanna2020open}. In many cases, a preprocessing step is required to process and prepare the data such that it is in a proper format for feature extraction and classification.

\vspace{3pt}
\noindent
\textbf{Preprocessing and Features.} 
When an RF signal is transmitted, the signal travels through a communication channel, which may have a negative effect (such as distortion) on the signal due to differences in the observed channel impulse response and additive noise. Therefore, various signal processing methods such as filtering can be applied to eliminate these effects ~\cite{shen2022towards, agadakos2020chameleons,hanna2020open,jian2020deep,jian2021radio} as well as reduce the effects of noise ~\cite {li2022reliable,lee2022aerokey,jian2020deep} or device movement~\cite{li2022reliable}.

The signal can be processed in different domains such as in the time domain, frequency domain, or both by using a Short-Time Fourier Transform (STFT) ~\cite{al2021deeplora}. In short, different signal processing methods can be employed to transform a received signal to the desired format.

Using the (transformed) input, fingerprinting approaches then focus on extracting features and characteristics useful towards securely authenticating devices.

Common features used in prior work are shown in Table~\ref{tab:fingerprinting}. As depicted in the table, there are many types of features such as peak frequency, frequency offset, signal-to-noise ratio (SNR), and statistical features \cite{hua2018accurate, givehchian2022evaluating}. Hardware imperfections also impart unique features onto the sent or measured signal which allows us to extract unique features for fingerprinting ~\cite{agadakos2020chameleons,al2020exposing,sankhe2019no,das2018deep,hanna2019deep, li2022reliable}. 

 \vspace{3pt}
\noindent
 \textbf{Classifier.}
 Many approaches are used in the RF space for classification. Deep learning is a prevalent approach for many modalities such as WiFi \cite{hanna2020open, jian2020deep, sankhe2019no, jian2021radio, al2020exposing}, ZigBee \cite{yu2019robust} and LoRa \cite{das2018deep}. Feature-based approaches such as support vector machine and kNN models are also used \cite{brik2008wireless, reising2020radio, joo2020hold}. Statistical approaches are also commonly used such as distance-based metrics ~\cite{givehchian2022evaluating}, profile analysis ~\cite{celosia2019fingerprinting}, and statistical t-tests ~\cite{chen2020eingerprint}.

 \vspace{3pt}
\noindent
 \textbf{Performance.}
 The detection range of RF fingerprinting approaches often depends on the modality, but was generally short. For instance, LoRa typically has a range in the thousands of meters but was tested in a 2240 square feet room in DeepLoRa \cite{al2021deeplora}. RFID, a short-range modality, understandably has a short range. Interestingly, in a work that fingerprinted drones, the range was short (around 8 meters), due to the distance limitations of mm-Wave signals ~\cite{li2022reliable}.

 The scalability of the systems varies as well. WiFi, BLE, and RFID are the most scalable, in that there are works that test thousands of devices \cite{jian2020deep, celosia2019fingerprinting, chen2020eingerprint}. However, the number of devices is typically tens to hundreds. One LoRa work tested 60 LoRa devices \cite{shen2022towards}, and many WiFi works tested in the hundreds of devices \cite{jian2021radio, restuccia2019deepradioid, brik2008wireless}.
Spoofing is considered in a few works, but not in many. Rogue devices are considered in a few LoRa and ZigBee works \cite{bihl2016feature, shen2022towards}. 

 \subsection{Network Fingerprinting}
 In network fingerprinting, the goal is to fingerprint devices based on a particular packet flow in the network. In contrast to RF, networking approaches focus on the analysis of network traffic compared to individual I/Q samples and packets.

\vspace{3pt}
\noindent
 \textbf{Input.}
 The input for network approaches is generally the observed traffic seen on a network. Each device sends a packet on the network which is then collected at the receiver as part of the packet flow that consists of multiple devices' output.
 
 \vspace{3pt}
\noindent
 \textbf{Preprocessing and Features.}
 In network approaches, typical preprocessing strategies include traffic parsing \cite{perdisci2020iotfinder} and the collection of data \cite{charyyev2020locality, sivanathan2018classifying}. Once data is preprocessed, we then isolate specific features, such as the timing of packets \cite{perdisci2020iotfinder}, clock offsets \cite{kohno2005remote}, RSS \cite{xu2015device}, and flow volume \cite{sivanathan2018classifying}. 
 
 \vspace{3pt}
\noindent
 \textbf{Classifier.}
 In network approaches, we typically see statistically-based or feature-based classifiers. Similarity metrics are often used as classifier models, such as in IoTFinder and many others ~\cite{perdisci2020iotfinder, miettinen2017iot, charyyev2020locality}. We see the usage of feature-based classifiers such as in one work that utilizes Naive Bayes and a random forest classifier \cite{sivanathan2018classifying}. 
 
 \vspace{3pt}
\noindent
 \textbf{Performance.}
 Generally speaking, the range of a network-based fingerprinter depends on the network the receiver is on. Thus, the range will generally be the range of the network.

 Network fingerprinting approaches generally consider devices in the region of tens of devices. Two approaches consider device counts in the twenties \cite{miettinen2017iot, sivanathan2018classifying}. Two other approaches considered 53 devices \cite{perdisci2020iotfinder} and 66 idle devices \cite{charyyev2020locality}.

 Most works do not consider the impacts of spoofing and adversarial attacks. However, a select few consider adversaries and rogue APs ~\cite{miettinen2017iot, xu2015device}.
 
 \subsection{Electromagnetic Fingerprinting}
Electromagnetic fingerprinting relies on measuring electromagnetic signals (EM) emitted from processors, memory, or hardware units. Oftentimes these signals are produced unintentionally via programmatic activity and have typically been used as a side-channel \cite{camurati2020understanding, nazari2017eddie, he2017hardware}. For fingerprinting, properties of these EM signals caused by uniqueness in the hardware itself can be used for authentication. 

\vspace{3pt}
\noindent
\textbf{Input.} In EM fingerprinting, inputs are usually EM signals that are emitted from different electronic devices, CPU, and memory activities. On the receiver end, the information is typically received as I/Q samples \cite{shen2022electromagnetic, feng2023fingerprinting}.

\vspace{3pt}
\noindent
\textbf{Preprocessing and Features.} Like other fingerprinting works, preprocessing is critical towards extracting features. Methods such as averaging ~\cite{lee2022aerokey} and filtering ~\cite{shen2022electromagnetic, feng2023fingerprinting} are typically used to reduce noise and prepare the signal for analysis \cite{lee2022aerokey}.

The next step after performing preprocessing is understanding the characteristics of the devices we are authenticating. Different pieces of hardware such as the CPU ~\cite{cheng2019demicpu, feng2023fingerprinting}, DRAM ~\cite{shen2022electromagnetic}, and integrated circuits ~\cite{ahmed2017radiated, ahmed2018authentication} emanate unique electromagnetic emanations per device. Generally speaking, electromagnetic fingerprinting approaches utilize the unique frequency response of circuits, processors, and memory components to identify between devices.

 \vspace{3pt}
\noindent
 \textbf{Classifier.}
 In EM Fingerprinting, a wide range of classifiers are used. Statistical approaches like cosine similarity are used in EM-ID and other approaches \cite{yang2016id, ahmed2018authentication, ahmed2017radiated}. Additionally, like RF fingerprinting, feature-based methods are a common strategy ~\cite{shen2022electromagnetic, cheng2019demicpu, feng2023fingerprinting}.

 \vspace{3pt}
\noindent
 \textbf{Performance.}
 Range typically falls short in electromagnetic approaches. Many methods utilize an electromagnetic probe to measure the signal, resulting in near-field ranges \cite{yang2016id, cheng2019demicpu, ahmed2018authentication, ahmed2017radiated}. Digitus ~\cite{feng2023fingerprinting} and MemScope ~\cite{shen2022electromagnetic} are exceptions with a range of around 10 meters and 30 meters, respectively, via leveraging CPU and memory clock signals.

 Scalability is typically in the tens of devices. On the short side, one approach only tests 5 same type devices \cite{mariano2019classification}. EM-ID tests 40 devices \cite{yang2016id}, MemScope tests 65 devices \cite{shen2022electromagnetic}, and DeMiCPU tests 90 devices (70 laptops and 20 phones) \cite{cheng2019demicpu}.

 Unfortunately, outside of Digitus and MemScope, none of the papers reviewed discussed spoofing or adversarial attacks in detail. Digitus considers spoofing and hidden devices ~\cite{feng2023fingerprinting}, whereas MemScope discusses the classification of alien devices as well as record-and-replay attacks \cite{shen2022electromagnetic}.

\vspace{3pt}
\begin{mdframed}[backgroundcolor=Apricot]
\vspace{-5pt}
\begin{lemm}\label{lem1}
\textbf{Current fingerprinting systems are modulation specific and often do not consider the presence of other devices, as well as adversarial scenarios.}
\end{lemm}
\end{mdframed}

%% file: Sections/hidden_devices.tex
Eavesdropping (hidden) device detection is important for securing IoT. Online or offline devices such as wireless cameras and microphones may be planted in a room by an adversary in order to collect sensitive information. Hidden device detection is similar to fingerprinting in that the system needs to identify a device. However, \textit{unlike fingerprinting}, hidden device detection systems need to be able to find (and localize) devices that have not been seen by the system, with localization block $L$ from Figure ~\ref{fig:formula_figure}. Thus, hidden device detection systems are often designed to identify devices via a particular modality (RF, Network, etc.) in order to aid detection. Additionally, these systems have their own processing blocks $p$ to extract new features to detect hidden devices which are sent into new classifiers $C$. In Table~\ref{tab:hidden_devices}, we display a subset of works that represent the various approaches seen in the literature. In the following subsections, we discuss the various approaches to hidden device detection for the various modalities. 

\begin{table*}[]
    \centering
    \caption{Relevant hidden device detection works. There are four classes of approaches: RF, Network, EM, and Thermal. Each method takes in a unique input. Some methods use more lightweight classifiers (feature-based and statistically-based), whereas others use more complex approaches (deep learning). These methods leverage unique characteristics of the device to distinguish between signals. Each method has different abilities in terms of range and detection time. Some approaches localize the device in the room, while others do not. Each system works for particular classes/kinds of transmitters and receivers. Finally, some methods are tested on a real setup and some are simulated.}
    \begin{tabular}{|>{\raggedright\arraybackslash}m{1.5cm}|>{\raggedright\arraybackslash}m{4cm}|>{\centering\arraybackslash}m{1cm}|>{\centering\arraybackslash}m{1cm}|>{\centering\arraybackslash}m{1cm}|>{\centering\arraybackslash}m{1cm}|c|c|c|>{\centering\arraybackslash}m{2cm}|>{\centering\arraybackslash}m{1cm}|>{\centering\arraybackslash}m{1cm}|}
    \hline
    \textbf{Paper}&\textbf{Notes}&\rotatebox{90}{\textbf{Modality}}&\rotatebox{90}{\textbf{Input}}&\rotatebox{90}{\textbf{Characteristic}}&\rotatebox{90}{\textbf{Classifier}}&\rotatebox[origin=c]{90}{\textbf{Range}}&\rotatebox[origin=c]{90}{\textbf{Time}}&\rotatebox[origin=c]{90}{\textbf{Localized?}}&\rotatebox{90}{\textbf{TX Device}}
    &\rotatebox{90}{\textbf{RX Device}}
    \\
    \hline

    Lumos \cite{sharma2022lumos} & Identify hidden WiFi-connected devices in a room via WiFi packets.
    & RF & Packet & Packet Features & F.B. & \fullcirc & \fullcirc & \fullcirc & Cameras, mics, doorbells, etc. & Macbook or RPI\\
    \hline

    Ghostbuster \cite{chaman2018ghostbuster} & Detect wireless eavesdroppers (passive receivers) via unintended RF leakage. & RF & Packet & Passive RF Leakage & Signal Processing & \halfcirc & \emptycirc & \emptycirc & USRP & USRP\\
    \hline

    E-Eye \cite{li2018eye} & Find hidden e-devices via mm-Wave sensing. & RF & mm-Wave Signal & Non Linear Effects & F.B. & \emptycirc & \emptycirc & \fullcirc & Laptop, tablet, wearable, etc. & mm-Wave Probe\\
    \hline

    Blink + Flicker \cite{liu2018detecting} & Detect spy cameras via changing ambient light and analyzing packet flow changes. & Network & Packet Flows & Packet Flow Changes & Stats. & \fullcirc & \halfcirc & \emptycirc & Wireless IP Cameras & Linux Tablet / RPI\\
    \hline
    
    SnoopDog \cite{singh2021always} & Discovering and localizing wireless sensors via user action and wireless traffic. & Network & Packet Flows & Traffic Patterns & Database & \halfcirc & \emptycirc & \fullcirc & Wireless Sensors & Mobile Phone\\
    \hline

    EarFisher \cite{shen2021earfisher} & Find eavesdroppers by analyzing their inherent memory EMR response. & EM & I/Q Samples & Memory EMR & Stats. & \fullcirc & N/A & \emptycirc & Laptop & BladeRF\\
    \hline

    HeatDeCam \cite{yu2022heatdecam} & Detect hidden spy cameras without wireless connectivity using thermal imagery. & Thermal & Thermal Input, Image & Heat Dissipation & D.L. & \halfcirc & \halfcirc & \fullcirc &  Spy Cameras & Mobile Phone\\
    \hline

    \end{tabular}
    \smallskip
    
    \textbf{Range} - \emptycirc \hspace{1mm}$<$ 1 m, \halfcirc \hspace{1mm}$<$ 5 m, \fullcirc \hspace{1mm}$>$ 5 m\\
    \textbf{Time} - \emptycirc \hspace{1mm}$<$ 1 min, \halfcirc \hspace{1mm}$<$ 5 min, \fullcirc \hspace{1mm}$>$ 5 min\\
    \textbf{Localized?} - \emptycirc \hspace{1mm}No, \fullcirc \hspace{1mm}Yes \\
    \label{tab:hidden_devices}
    
\end{table*}

\subsection{RF Approaches}\label{subsection:rf}
RF hidden device detection attempts to discover a hidden device via either their packet features \cite{sharma2022lumos} or a device's unique response \cite{chaman2018ghostbuster, li2018eye}. In the first case, we need to listen to the normal activity of the device. In the second case, by sending a specific input, we can elicit a certain detectable response from the device in order to detect it. In the following, more details about the various steps in this process are provided. 

\vspace{3pt}
\noindent
\textbf{Input.} For RF devices, hidden device detection depends on the transmission of packets via transmitter modules and the resulting packets or I/Q samples that the detector receives. For some approaches, the packets sent from the hidden device are used to detect the device itself. Many works utilize WiFi packets in their detection system \cite{sharma2022lumos, chaman2018ghostbuster, li2018adversarial, nguyen2017matthan}. Other works depend on a stimulus or an input that invokes a response from the device. These approaches utilize the raw I/Q samples from the device \cite{li2018eye, park2006hidden, stagner2011practical, stagner2012locating, wild2005detecting}. In either case, the detectors receive the data as input, conduct any preprocessing (averaging, smoothing, etc.) needed to prepare the signal for classification, then send the preprocessed data to the detector.

\vspace{3pt}
\noindent
\textbf{Features.} The features used to detect a device depend on the received input. In many cases, receiver leakage or characteristics of unintended emissions can be used as features to detect a device \cite{li2018eye, chaman2018ghostbuster, park2006hidden, stagner2011practical, stagner2012locating, wild2005detecting, nguyen2017matthan}. In other cases, features from the packet itself, such as RSS information, can be used \cite{sharma2022lumos, li2018adversarial}. 

\vspace{3pt}
\noindent
\textbf{Classifiers.} How the above features are used depends on the classifier involved in the detection. Various kinds of classifiers are used to detect the presence of hidden devices, ranging from lightweight statistical approaches to complex machine learning approaches.
Many approaches utilize signal processing and statistics to identify hidden devices ~\cite{nguyen2017matthan, park2006hidden}. Some common techniques include filtering, sampling, applying transformations, and subtracting out unwanted signals ~\cite{chaman2018ghostbuster, stagner2011practical, stagner2012locating}.

Some approaches utilize a feature-based approach. A feature-based approach depends on identifying a unique response from a device. This unique response could be features analyzed from a packet ~\cite{sharma2022lumos}, or from features derived from device measurements ~\cite{li2018eye}. These features can be then used as inputs into lightweight models such as XGBoost and support vector machine ~\cite{sharma2022lumos, li2018eye}.
Beyond these approaches, one can utilize RSS as input for localization algorithms \cite{li2018adversarial}. RSS information can be received within the packet and be used as input to a log-distance path loss model for instance.

\vspace{3pt}
\noindent
\textbf{Performance.} To gauge performance, we need to analyze the devices used, the detection range and time, and whether the approach localizes the device or not. These are all factors that make a hidden device detection system feasible or not. 

For the receiving device, a common device used is a software-defined radio \cite{chaman2018ghostbuster, stagner2011practical, stagner2012locating, nguyen2017matthan}. Laptops or smartphones are often used as well \cite{sharma2022lumos, li2018adversarial}. These devices are often used for wireless reception.

A variety of hidden devices are tested including other laptops \cite{li2018eye}, cameras \cite{sharma2022lumos, li2018adversarial}, microphones \cite{sharma2022lumos}, and GMRS radios \cite{stagner2011practical, stagner2012locating}. 

The range of detection varies depending on the technique and technology. Detection can be near-field in mm-Wave applications ~\cite{li2018eye} or within the space of a room such as in Lumos \cite{sharma2022lumos} and Ghostbuster \cite{chaman2018ghostbuster}. Finally, approaches such as Matthan can achieve tens to hundreds of meters \cite{stagner2012locating, nguyen2017matthan}.

The \textit{time} of detection varies as well. Some detect under a minute \cite{chaman2018ghostbuster, wild2005detecting} while others take many minutes \cite{sharma2022lumos, li2018adversarial}. A shorter detection time is better because it reduces the time the user could potentially be exposed to an adversary.

Finally, only a few approaches \textit{localize} the device. In Lumos \cite{sharma2022lumos}, the device is localized within 1.5 meters. Two other approaches localize the device under 5 meters \cite{stagner2012locating, li2018adversarial}.

\subsection{Network Approaches}\label{subsection:net} 
Network hidden device works are mostly similar to RF hidden device works. However, they vary from RF due to utilizing packet flows and analyzing network traffic compared to I/Q samples or individual packets.

\vspace{3pt}
\noindent
\textbf{Input.} For network-based approaches, the input is generally simple. Given a network and a capturing device, the packet flows can be gathered. These packet flows can then be parsed and analyzed for the sake of hidden device detection.

\vspace{3pt}
\noindent
\textbf{Features.} The features utilized vary depending on the work. Some works gather unique network information from the received packets \cite{saidi2020haystack, siby2017iotscanner}, while other works analyze changes in network traffic caused by devices \cite{singh2021always, heo2022there, ortiz2019devicemien, liu2018detecting, cheng2018dewicam, wu2019you, he2021motioncompass}. CSI:DeSpy captures WiFi camera video bitrate fluctuations as well as CSI data \cite{salman2022csi}.

\vspace{3pt}
\noindent
\textbf{Classifiers.} Network-based approaches use a wide range of classifiers to detect hidden devices.

Mapping (characteristic to a device) and rule-based classifiers can be used, where a dictionary or database is used to store device mappings and data of devices ~\cite{saidi2020haystack, singh2021always}. Signal processing and statistics techniques such as correlation, peak-finding, statistical testing, and thresholding can be used to detect devices ~\cite{heo2022there, liu2018detecting}.

Machine learning algorithms can be used to detect hidden devices, ranging from lightweight feature-based classifiers ~\cite{cheng2018dewicam, salman2022csi, ortiz2019devicemien} to more complex deep learning methods ~\cite{wu2019you}.

\vspace{3pt}
\noindent
\textbf{Performance.} As this approach utilizes network traffic, there are some built-in specifications for performance.

The receiver (network traffic analyzer) can be implemented on many devices such as smartphones \cite{singh2021always, salman2022csi, heo2022there, cheng2018dewicam}, laptops \cite{siby2017iotscanner}, Raspberry Pis \cite{liu2018detecting}, or access points \cite{ortiz2019devicemien}.

Many transmitting devices are tested such as wireless cameras \cite{siby2017iotscanner, salman2022csi, heo2022there, he2021motioncompass, liu2018detecting, cheng2018dewicam, wu2019you}. In general, many kinds of IoT devices beyond cameras can be tested \cite{saidi2020haystack, ortiz2019devicemien}, as long as the device has a wireless connection. This includes devices such as doorbells, bulbs, and smart speakers.

The range for network-based approaches is typically as large as the network range. Only a few approaches are limited in range due to reliance on motion input. For example, for SnoopDog \cite{singh2021always}, the user must be $<$ 5 meters from the device. For MotionCompass \cite{he2021motioncompass}, the user must be within the motion detection range of the camera.

The time of detection varies. Some are within a minute \cite{salman2022csi, cheng2018dewicam, wu2019you}, while others can be multiple minutes \cite{heo2022there, liu2018detecting} to an hour \cite{saidi2020haystack}. Similarly to RF approaches, a shorter detection time is better to limit the negative consequences on users.

A few works consider localization. SnoopDog \cite{singh2021always} can localize within a small region of the room. Works such as MotionCompass can localize a device within a couple of centimeters \cite{he2021motioncompass, heo2022there}.

\subsection{Electromagnetic Approaches}\label{subsection:em} 

Electromagnetic approaches utilize electromagnetic radiation (EMR) from devices in order to detect. We analyze three works in this section. In EarFisher \cite{shen2021earfisher}, the source of these emissions are from the memory unit. In DeHiRec \cite{zhou2022dehirec}, the emissions are from the ADC. Another work, by Thotla \cite{thotla2013detection} utilizes emissions from a super-regenerative receiver (SRR).

\vspace{3pt}
\noindent
\textbf{Input.} In all three works, the received input is I/Q samples. The EMR is radiated at a certain frequency depending on the device. For instance, in EarFisher, the frequency of the memory unit itself dictates where the EMR is seen in the spectrum. Thus, the detector needs to find the location of the EMR in the spectrum and capture the signal.

\vspace{3pt}
\noindent
\textbf{Features.} In all three cases, the detector attempts to collect unintentional emissions from the hidden device. In EarFisher, the adversary sends bait traffic into the air. The device to be detected then receives the message and stores data in memory, which produces a measurable EMR response. In DeHiREC, the authors utilize an EMR catalyzing method to produce EMR from the ADC. In Thotla's work, the authors stimulate the SRR with a sinusoidal signal.

\vspace{3pt}
\noindent
\textbf{Classifiers.} All three works utilize stats and signal processing to detect devices. EarFisher uses statistical hypothesis testing for detection purposes. DeHiRec compares the EMR to the stimulation. Thotla's work uses thresholding and probability,

\vspace{3pt}
\noindent
\textbf{Performance.} The receiving devices vary from a BladeRF SDR \cite{shen2021earfisher} to an oscilloscope \cite{thotla2013detection}. The transmitters consist of devices such as offline microphones \cite{zhou2022dehirec}, laptops \cite{shen2021earfisher} and doorbells \cite{thotla2013detection}.

The range of detection varies. DeHiREC requires a close range (0.2 meters). EarFisher and Thotla's works have a range of up to 30 meters and 26 meters, respectively, demonstrating the potential to detect devices at range using EMR.
Time and localization information were not discussed much. For DeHiREC, the device is localized due to the short range. For EarFisher and Thotla's work, localization was not discussed.

\subsection{Thermal Approaches}\label{subsection:thermal}
Thermal approaches attempt to discover a device by detecting the heat dissipation from a device. We present two works: HeatDeCam \cite{yu2022heatdecam} and See No Evil \cite{zuniga2022see}. In these two works, the devices in question are hidden cameras.

\vspace{3pt}
\noindent
\textbf{Input.} In See No Evil, the authors utilize thermal images as input. In HeatDeCam, the authors augment the thermal input received by combining it with a masked visual input (an image). 

\vspace{3pt}
\noindent
\textbf{Features.} The key insight is to leverage the heat dissipation from a device. The location of the camera will look "warmer" than the rest of the room, indicating a device. 

\vspace{3pt}
\noindent
\textbf{Classifiers.} HeatDeCam utilizes deep learning (specifically a ResNet architecture) to identify the hidden device. See No Evil utilizes image preprocessing and thresholding to clean the image and separate the device from the background.

\vspace{3pt}
\noindent
\textbf{Performance.} Both HeatDeCam and See No Evil utilize smartphones as their receiving device. Additionally, the target of each system is hidden cameras.

As both systems detect devices via their thermal images, the range is naturally limited. The range of See No Evil and HeatDeCam are several meters. Both systems also take similar time durations to detect. The user needs to take pictures around the room, then classify the pictures. This takes several minutes.

Finally, as both these approaches use thermal imagery, the device is localized within a general region.

\begin{table*}[]
    \centering
    \caption{Relevant spoofing and adversarial attacks/defenses for various modalities, which dictates the characteristic(s) used. Some attacks require a classifier, while others do not. In this space, there are both attacks and defenses, and these can be either targeted or untargeted. Each approach has a certain range and time of attack/defense, with certain TX and RX devices for each approach. Finally, some approaches are tested on a physical setup whereas others are simulated.}
    \begin{tabular}{|>{\raggedright\arraybackslash}m{1.5cm}|>{\raggedright\arraybackslash}m{3cm}|>{\centering\arraybackslash}m{1cm}|>{\centering\arraybackslash}m{1cm}|>{\centering\arraybackslash}m{1cm}|>{\centering\arraybackslash}m{1cm}|c|c|c|c|>{\centering\arraybackslash}m{1cm}|>{\centering\arraybackslash}m{1cm}|>{\centering\arraybackslash}m{1cm}|>{\centering\arraybackslash}m{1cm}|>{\centering\arraybackslash}m{1cm}|}
    \hline
    \textbf{Paper}&\textbf{Notes}&\rotatebox{90}{\textbf{Modality}}&\rotatebox{90}{\textbf{Input}}&\rotatebox{90}{\textbf{Characteristic}}&\rotatebox{90}{\textbf{Classifier}}&\rotatebox[origin=c]{90}{\textbf{Attack/Defense}}&\rotatebox[origin=c]{90}{\textbf{Targeted/Untargeted}}&\rotatebox[origin=c]{90}{\textbf{Range}}&\rotatebox[origin=c]{90}{\textbf{Time}}&\rotatebox{90}{\textbf{TX Device}}
    &\rotatebox{90}{\textbf{RX Device}}
    &\rotatebox{90}{\textbf{Setup}}
    \\
    \hline

    BIAS \cite{antonioli2020bias} & Impersonate a BLE device during secure connection establishment. & RF & Packet & Address, Role Switch & N/A & \fullcirc & \fullcirc & \halfcirc & \emptycirc & Eval. Board & BLE Device & \fullcirc \\
    \hline

    RFAL \cite{roy2019rfal} & Identify rogue transmitters via a GAN. & RF & I/Q Samples & I/Q Perturbations & D.L. & \emptycirc & \emptycirc & \halfcirc & \fullcirc & USRP & RTL-SDR & \fullcirc \\
    \hline

    GAN Attack \cite{shi2020generative} & Utilize a GAN to produce spoofed signals to fool a fingerprinter. & RF & I/Q Samples & I/Q Perturbations & D.L. & \fullcirc & \fullcirc & N/A & \fullcirc & N/A & N/A & \emptycirc \\
    \hline
    
    GAN Attack \cite{karunaratne2021penetrating} & Fool RF authentication by adding perturbations in the sent signal. & RF & Rx Feedback & I/Q Perturbations & Markov Decision Process & \fullcirc & \emptycirc & \fullcirc & \fullcirc & ADALM Pluto SDR & ADALM Pluto SDR & \halfcirc \\
    \hline

    CNN Attack \cite{bao2021threat} & Applies traditional adversarial attack techniques towards device identification. & RF & I/Q Feature Map & I/Q Perturbations & D.L. & \fullcirc & \halfcirc & N/A & \fullcirc & Signal Generator & Signal Receiver & \emptycirc \\
    \hline

    SemperFi \cite{sathaye2021semperfi} & Minimize the effect of adversarial GPS spoofing. & GPS & GPS Signal & Adv. Peak & Stats. & \emptycirc & \emptycirc & \fullcirc & \emptycirc & Drone & SDR & \halfcirc \\
    \hline

    Hu-Fu \cite{wang2018towards} & RFID authenticator that is resistant to spoofing and replay attacks. & RFID & RFID Signal & Inductive Coupling & N/A & \emptycirc & \fullcirc & \emptycirc & \emptycirc & RFID & USRP & \fullcirc \\
    \hline

    Blinding Attack ~\cite{oconnor2019blinded} & Blind and confuse actuators to disrupt IoT operation. & Network & Packet & Division Telemetry & N/A & \fullcirc & \fullcirc & \fullcirc & \halfcirc & Wireless Router & Home Devices & \fullcirc \\
    \hline

    \end{tabular}
    \smallskip

    \textbf{Attack/Defense} - \emptycirc \hspace{1mm}Defense, \halfcirc \hspace{1mm}Attack and Defense, \fullcirc \hspace{1mm}Attack \\
    \textbf{Targeted/Untargeted} - \emptycirc \hspace{1mm}Untargeted, \halfcirc \hspace{1mm}Targeted and Untargeted, \fullcirc \hspace{1mm}Targeted \\
    \textbf{Range} - \emptycirc \hspace{1mm}$<$ 1 m, \halfcirc \hspace{1mm}$<$ 5 m, \fullcirc \hspace{1mm}$>$ 5 m\\
    \textbf{Time} - \emptycirc \hspace{1mm}$<$ 1 min, \halfcirc \hspace{1mm}$<$ 5 min, \fullcirc \hspace{1mm}$>$ 5 min\\
    \textbf{Setup} - \emptycirc \hspace{1mm}Simulated, \halfcirc \hspace{1mm} Simulated and Real Physical Setup, \fullcirc \hspace{1mm}Real Physical Setup\\
    
    \label{tab:spoofing}
\end{table*}

\vspace{3pt}
\begin{mdframed}[backgroundcolor=Apricot]
\vspace{-5pt}
\begin{lemm}\label{lem1}
\textbf{Many eavesdropping works do not actually localize the device and require close access to the device.}
\end{lemm}
\end{mdframed}

%% file: Sections/spoofing.tex
When developing a fingerprinting system, it is important to consider the impact of adversaries on the system, who have their own processing steps ($\Sigma$, $p$ in Figure~\ref{fig:formula_figure}) and $Adversary$ $Models$ to fool the fingerprinter. For example, an attacker can spoof the device. Spoofing is when the adversary attempts to fool the classifier into thinking they are a specific authenticated device (a form of targeted attack), and/or protect themselves from being detected (i.e., a hidden device).  An attacker can also conduct an adversarial attack in which the attacker adds perturbations to their signal to fool the classifier, either targeted or untargeted. In either case, the integrity of the fingerprinting system can be put in jeopardy and the system must be designed with security in mind. In the following subsections, we discuss various attacks and defenses for fingerprinting systems. A selection of various works that cover the existing methodologies of spoofing and adversarial attacks are shown in Table~\ref{tab:spoofing}. 

\subsection{Attacks}\label{subsection:attacks}

\vspace{3pt}
\noindent
\textbf{Input.} In most cases, the adversary leverages raw I/Q samples to attack \cite{bao2021threat, shi2020generative, xu2022colluding, bahramali2021robust, merchant2019securing, xu2023adversarial}. For BLE spoofing, which involves the spoofing of BLE device identity when pairing, BLE packets are used \cite{ai2022blacktooth, antonioli2020bias}. In Karunaratne's work, the attacker receives feedback from the receiver as input to generate adversarial samples \cite{karunaratne2021penetrating}.

\vspace{3pt}
\noindent
\textbf{Method of Attack.} There are a few methods of attack. It is common for receiver feedback to be used in tandem with a model ~\cite{karunaratne2021penetrating}. One specific model being used is a generative adversarial network (GAN) \cite{shi2020generative, xu2022colluding, merchant2019securing, bahramali2021robust}, which can be used to generate spoofed samples in either a targeted or untargeted setting. In the case of BLE, a model is not used but rather the direct spoofing of identity \cite{ai2022blacktooth, antonioli2020bias}. Finally, some works have analyzed traditional methods in adversarial machine learning such as FGSM, BIM, and PGD to launch both targeted and untargeted attacks \cite{bao2021threat, xu2023adversarial}.

For these methods discussed above, the general goal is to create perturbations in the sent signal to fool the fingerprinting classifier.

Another class of attacks attempts to "blind" the fingerprinter. These works attempt to disable sensors or actuators ~\cite{oconnor2019blinded} or disable or hinder the IoT device ~\cite{fu2022iot}.

\vspace{3pt}
\noindent
\textbf{Fingerprinting Model Attacked.} Attacks generally consider attacking deep learning models \cite{bao2021threat, shi2020generative, xu2022colluding, bahramali2021robust, merchant2019securing, xu2023adversarial}. Karunaratne attacks a discriminator and an LSTM system \cite{karunaratne2021penetrating}.

\vspace{3pt}
\noindent
\textbf{Performance.} In gauging the performance of a spoofing or adversarial attack, we need to consider many factors. First, are the experiments simulated or conducted in a real testbed setting? Second, what is the range of this setup? Third, how long does the attack take? Finally, is physical access needed to the devices?

Most of the attacks are simulated or at least conducted partially in a simulated setting. The attacks that leverage traditional methods in adversarial machine learning generally do not consider over-the-air methods and only consider attacking the received input \cite{bao2021threat, xu2023adversarial}. 

Since many of the attacks were simulated, those works do not mention the distance that this attack was successful. For the works that are conducted in real settings, the range of the wireless medium is assumed \cite{ai2022blacktooth, antonioli2020bias, karunaratne2021penetrating} or within the wireless range ~\cite{oconnor2019blinded, fu2022iot}.

The time to conduct the attack is generally tied to the time it takes to train the model \cite{bao2021threat, shi2020generative, xu2022colluding}. In the case of BLE attacks, the attack can be as quick as seconds or within pairing time \cite{ai2022blacktooth, antonioli2020bias}.

In many cases, physical access to the devices is needed in order to place an adversarial receiver close to the receiver \cite{shi2020generative} or to access the samples for adversarial machine learning techniques \cite{bao2021threat, xu2023adversarial}. In some works, the adversary needs to be close enough to the victim devices in order to capture their signals \cite{xu2022colluding, ai2022blacktooth, antonioli2020bias} or the receiver's signals \cite{karunaratne2021penetrating}.

\subsection{Defenses}\label{subsection:defenses}

\vspace{3pt}
\noindent
\textbf{Input.} Defenses towards spoofing or adversarial attacks receive the same kind of signal (I/Q samples, packets, etc.), no matter if the signal is trusted or adversarial. I/Q samples are used in many works \cite{bahramali2021robust, merchant2019securing, xu2023adversarial, roy2019rfal}. Features of the packet itself can be used as well, such as CSI and RSS \cite{liu2019real, chen2007detecting, sheng2008detecting}. In other cases, the raw signal itself is used, such as in GPS signals \cite{sathaye2021semperfi} or RFID signals \cite{wang2018towards}.

\vspace{3pt}
\noindent
\textbf{Method of Defense.}
The prescribed defense for spoofing and adversarial attacks depends on the signal being sent and the motivation of the defender. For example, a defender can attempt to prevent attacks from happening in the first place. By using techniques such as adversarial training ~\cite{merchant2019securing, bahramali2021robust, xu2023adversarial} or perturbation subtracting \cite{bahramali2021robust}, in addition to designing the authentication scheme more securely ~\cite{wang2018towards}, a defender can train the classifier to be robust against adversarial samples.

A defender can also design a defense to detect adversaries. Anomaly detectors can be designed to separate authenticated devices from foreign devices. These can be designed by analyzing deviations ~\cite{sathaye2021semperfi} or by designing more complex models ~\cite{chen2007detecting, sheng2008detecting, roy2019rfal}.

\vspace{3pt}
\noindent
\textbf{Performance.} Similarly to attacks, we need to gauge the performance of a defense. Are the experiments conducted in a simulated or real setting? What is the range between the detector and the adversary? How long does detection take?

In most cases, the setup exists in a physical testbed \cite{wang2018towards, liu2019real, roy2019rfal, chen2007detecting, sheng2008detecting}. The distance between the detector and the adversary is generally large. The range can be as small as 10 feet \cite{roy2019rfal} or as large as 4.1 km \cite{sathaye2021semperfi}. Works that leverage RSS have a range in the tens of meters \cite{chen2007detecting, sheng2008detecting}.

The time of defense is generally quick and can be a few seconds \cite{sathaye2021semperfi, wang2018towards, liu2019real} or a few minutes \cite{chen2007detecting, sheng2008detecting}.

\begin{table*}[]
    \centering
    \caption{Attacks can be designed with many assumptions. i) Attack Type. The victim can be attacked various ways and the adversary can have different levels of model access. ii) Trusted Rx. The adversary can have varying access to the fingerprinter and database. iii) Environment. Various assumptions can be made on the channel as well as the presence of trusted Tx's.}
    \begin{tabular}{|>{\raggedright\arraybackslash}m{2cm}|>{\raggedright\arraybackslash}m{2.5cm}|>{\raggedright\arraybackslash}m{7cm}|}
    \hline
    \multicolumn{2}{|c|}{\textbf{Assumptions}}&\textbf{Values}
    \\
    \hline
    \multirow{2}{*}{\textbf{1)} Attack Type} & \textbf{a)} Victim & \textbf{x)} Untargeted \hspace{0.5cm}\textbf{y)} Targeted \hspace{0.86cm}\textbf{z)} Blinding \\
    \cline{2-3}
     & \textbf{b)} Model & \textbf{x)} White Box \hspace{0.56cm}\textbf{y)} Black Box \\
    \cline{2-3}
    \hline

    \multirow{2}{*}{\textbf{2)} Trusted Rx} & \textbf{a)} Physical Access & \textbf{x)} Access \hspace{1.07cm}\textbf{y)} No Access \\
    \cline{2-3}
     & \textbf{b)} Database & \textbf{x)} Access \hspace{1.07cm}\textbf{y)} No Access \\
    \cline{2-3}
    \hline

    \multirow{3}{*}{\textbf{3)} Environment} & \textbf{a)} Channel (train) & \textbf{x)} No Channel \hspace{0.4cm}\textbf{y)} Sim Channel \hspace{0.35cm}\textbf{z)} Real Channel \\
    \cline{2-3}
     & \textbf{b)} Channel (test) & \textbf{x)} No Channel \hspace{0.4cm}\textbf{y)} Sim Channel \hspace{0.35cm}\textbf{z)} Real Channel \\
    \cline{2-3}
    & \textbf{c)} Trusted Tx & \textbf{x)} Not Present \hspace{0.42cm}\textbf{y)} Present \\
    \cline{2-3}
    & \textbf{d)} Interference & \textbf{x)} Not Present \hspace{0.42cm}\textbf{y)} Multi Sources \\
    \cline{2-3}
    \hline

    \end{tabular}
    \smallskip
    
    \label{tab:attack}
    
\end{table*}

\vspace{3pt}
\begin{mdframed}[backgroundcolor=Apricot]
\vspace{-5pt}
\begin{lemm}\label{lem1}
\textbf{Attacks in existing works are largely unrealistic. They mostly only use a simulated environment, and require much access to the fingerprinting model.}
\end{lemm}
\end{mdframed}

%% file: Sections/attack.tex
\begin{figure}[t]
    \centering
    \includegraphics[width=0.9\columnwidth]{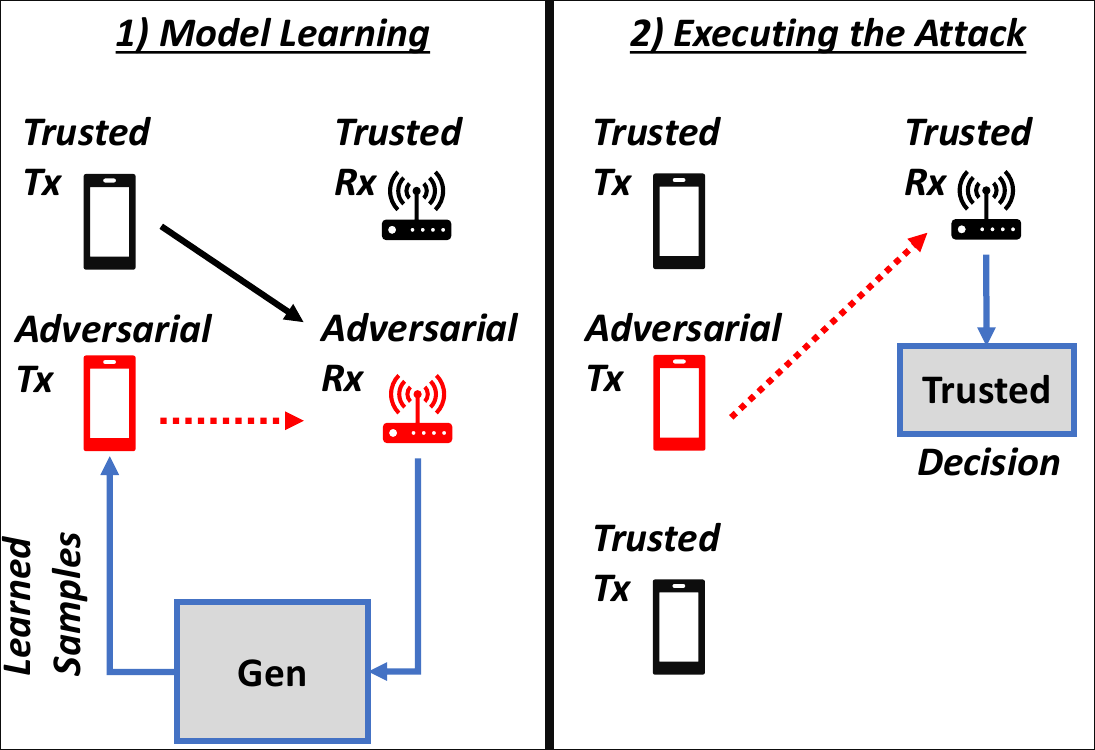}
    \caption{
    1) The adversary places their Rx near the trusted Rx and receives samples from the trusted Tx and the adversarial Tx. These samples are sent to a generative model (Gen), which iteratively learns how to generate samples that mimic the trusted Tx. 2) The adversarial Tx sends samples over the air to the trusted Rx, in the presence of trusted Tx's, and fools the trusted Rx.}
    \label{fig:attack_scenario}
\end{figure}

In this section, we provide a sample attack scenario that builds off existing fingerprinting approaches and leverages the improvements discussed in Section ~\ref{sec:disc} as well as the fingerprinting model discussed in Section ~\ref{sec:back} to devise a more robust attack. Table ~\ref{tab:attack} outlines the various assumptions one can make in an attack, and for this attack, we assume a set of assumptions of: \textbf{\{1ay, 1by, 2ay, 2by, 3az, 3bz, 3cy\}}, where \textbf{1ay} refers to a targeted attack type. Figure ~\ref{fig:attack_scenario} shows the experimental setup for our adversary. By outlining this attack, we hope to demonstrate a methodology that future attacks and defenses towards fingerprinting systems can follow.

\subsection{Attack Setup}\label{subsection:scenario}
To begin the attack, an adversary needs to develop samples that will fool an authenticator (assuming no physical access to the trusted Rx and database). This authenticator may have a specific modulation (or not). Ideally, the attacker wants to be agnostic to the modulation thus they only collect samples at the relevant frequencies. To minimize the impact of channel differences on data collection, the adversary ideally would like to collect samples by placing a receiver close to the trusted fingerprinter. Otherwise, the attack has to deal with variations in the channel (3a and 3b). Additionally, multiple trusted transmitters may be active at the same time creating further interferences (3d).

Note that these are important assumptions that are mainly lacking in prior work described in Section~\ref{sec:spoof}. 

\subsection{Model Learning}\label{subsection:learning}
An adversary can use different models. As explained earlier, with the advancement in machine learning, generative models (LLMs, GANs, Diffusion, etc.) are the ideal candidates for model learning. For example, using a generator and a discriminator, the attacker will iteratively receive samples, pass them to a GAN, and fine-tune the model to generate adversarial samples to attack the fingerprinter. The adversary can use preprocessing, filtering, and other iterative processing techniques to improve the quality of the spoofed samples. Furthermore, the model building can be improved if the adversary has white-box access to the fingerprinting model and/or can query the model to improve the generated samples.

\subsection{The Attack}\label{subsection:attack}
To conduct an attack, the learned samples should be transmitted over the air so that the trusted fingerprinter can observe. As explained in Section~\ref{sec:spoof}, prior work mostly considered a simplified version where the environment was simulated. An important thing to note, however, is that the robustness of the attack is dependent on the location of the transmitter and the channel. As a result, the robustness of attack should be examined when the transmitter is placed at various distances and paths from the trusted fingerprinter, in the presence of other trusted transmitters. Furthermore, the model can be fine-tuned to account for this variable.

%% file: Sections/discussions.tex
\subsection{Improving the Fingerprinting Model}\label{subsection:disc_improve}
One limitation of existing works is that fingerprinting approaches are often modulation dependent, as seen in Table~\ref{tab:fingerprinting}. For instance, a fingerprinter for WiFi devices may not work for ZigBee Devices. Additionally, hidden device detectors that utilize WiFi packet information may not leverage electromagnetic radiation information. Thus, fingerprinting schemes that can authenticate devices under different modalities (e.g., modulation agnostic) and leverage multiple sources of information is an area of future work.

Another limitation is many hidden device works do not consider the localization of the device. While just detecting the device is useful, an end-to-end pipeline that first 1) detects the device, and then 2) localizes the device would be \textit{more} useful.

\subsection{Designing the Experiments}\label{subsection:disc_experiment}
One limitation of existing works is many experiments are simulated (e.g., not using real data), especially in the adversarial domain as seen in Table~\ref{tab:spoofing}. Many of these works lack the final step of the attack (sending samples over the air) or do not consider the effects of outside interference. Conducting real experiments using a physical testbed and sending data over the air would demonstrate greater feasibility for the attacks.

Additionally, when designing an attack or creating a hidden device detector, physical access should also be taken into account. In a lot of works, the adversary needs to have access to the victim's device(s) or to be near it. Similarly, many hidden device works require the user to be directly adjacent to the device. In future works, limiting access and increasing the range of these systems would bolster these attacks or systems.

\subsection{Diversifying the Scenarios}\label{subsection:disc_improve}

Another limitation is many fingerprinting works do not consider the presence of other similar or dissimilar devices in the room active at the same time. When a fingerprinting model is deployed, the device cannot be physically isolated all the time. Thus, scenarios with multiple devices active at the same time should be considered and our models need to be designed with this scenario in mind.

Another limitation of existing works is that spoofing and adversarial scenarios are not always considered when designing a fingerprinting system. To demonstrate a system's robustness, we need to consider the presence of adversaries who hope to gain unauthorized access. Thus, future work should consider adversarial scenarios when developing fingerprinting systems.

%% file: Sections/conclusions.tex
We investigated current trends in device fingerprinting amid the onset of advanced machine learning techniques in order to improve the security of future IoT fingerprinting schemes. We also presented a fingerprinting model that offers insights into factors that influence fingerprinting and how an adversary can exploit fingerprinting weaknesses. To advance work in this field, we believe researchers should:

\begin{itemize}[leftmargin=5.5mm]
    \item Consider a more intricate fingerprinting model.
    \item Design experiments to test real-world applicability.
    \item Test a variety of scenarios, including adversarial settings.
\end{itemize}

By considering these points in future fingerprinting designs, we can increase the trust and reliability of IoT.